\documentclass[twocolumn,showpacs,preprintnumbers,amsmath,amssymb,aps,prb,superscriptaddress]{revtex4-2}
\usepackage[english]{babel}
\usepackage[utf8]{inputenc}
\usepackage{physics}
\usepackage{amstext}
\usepackage{bbold}
\usepackage[colorinlistoftodos, color=red!40, prependcaption]{todonotes}
\usepackage{cleveref}
\newcommand{\ceil}[1]{\lceil {#1} \rceil}
 
\begin{document}
\title{Quantum transfer in high-root topological insulators}
    
\author{G. F. Moreira}
\affiliation{Department of Physics $\&$ i3N, University of Aveiro, 3810-193 Aveiro, Portugal}

\author{A. Lykholat}
\affiliation{Department of Physics $\&$ i3N, University of Aveiro, 3810-193 Aveiro, Portugal}

\author{R. G. Dias}
\affiliation{Department of Physics $\&$ i3N, University of Aveiro, 3810-193 Aveiro, Portugal}

\author{A. M. Marques}
\email{anselmomagalhaes@ua.pt}
\affiliation{Department of Physics $\&$ i3N, University of Aveiro, 3810-193 Aveiro, Portugal}

\begin{abstract}
This paper focuses on the quantum state transfer in a one-dimensional (1D) high-root topological insulator (HRTI) with an arbitrary number of domains. We present the possibility of having multiple transfer processes in the same model due to the existence of various edge states in distinct energy gaps, which may benefit recent (de)multiplexing technologies.
We also derived the relations between transfer times of different root models and different gaps in the same model. We show how the exponential decay in transfer time caused by the fragmentation of a parent chain into domains can be generalized to its higher-root versions while maintaining a high transfer fidelity, and how the increasing number of domain wall states leads to a higher transfer fidelity against a general disorder regime due to the topological protection inherited from the parent model.
\end{abstract}

\maketitle

\section{Introduction} \label{sec:section1}

Topological insulators \cite{Asboth} (TIs) have been an intensely studied topic in condensed matter physics over the last decade due to their insulating bulk and topologically protected boundary states \cite{disorderplatero}. Recent studies \cite{apollaro201299,dlaska2017robust,lang2017topological,mei2018robust,d2020fast} show how these surface states can be used to transport information, particularly in 1D systems where fragmentation into multiple domains \cite{domainwalls} expedites the transferring process \cite{platero}. 
The exponential speed up of the state transfer enhances the robustness of the protocol against certain types of disorder and increases its coherence time, proving to be beneficial for applications in quantum computation. 
This approach sparked the interest of coupling the Su-Schrieffer-Heeger (SSH) model (a one-dimensional bipartite tight-binding model) with: a central atom that allows the transfer of its excitations to both ends of the whole chain \cite{wang2023realizing,zhang2024topological}, an embedded quantum ring that can construct a
channel with tunable output ports for quantum state transfer \cite{wang2024adiabatic}, and a superconducting resonator chain that behaves as a synthetic one-domain SSH model with a fully tunable single particle transport \cite{hanefficient}.

An interesting recent development in the field of topology came with the discovery of square-root TIs $(\sqrt{\text{TIs}})$ \cite{arkinstall2017topological} as physical systems whose Hamiltonian matrix representation in the Wannier basis becomes a block diagonal matrix after squaring it.
Subsequent systematic construction of $\sqrt{\text{TIs}}$ \cite{PhysRevResearch.2.033397} paved the way for higher-order topology systems, namely $2^n$-root TIs $(\sqrt[2^n]{\text{TIs}})$ \cite{PhysRevB.103.235425,marques20212}. These models present multiple band gaps with the possibility of displaying edge states in each of them and hence can be used in quantum state transfer protocols due to their inherited topological protection against certain disorder regimes \cite{marques20212,viedma2024topological,zurita2025hidden}. 
One of these $\sqrt[2^n]{\text{TIs}}$ is the sine-cosine model of degree $n$ [SC($n$)], a chain with $2n$ sites per unit cell and hopping parameters given by sinusoidal functions of a set of angles $(\sin\theta_1^{(n)},\cos\theta_1^{(n)},\sin\theta_2^{(n)},\cos\theta_2^{(n)},...,\sin\theta_n^{(n)},\cos\theta_n^{(n)})$. This model has a unique property of being its own root model, i.e., for a specific set of angles, one can square the Hamiltonian of a SC($2n$) chain, apply an energy downshift and hopping unit renormalisation $n$ times, iteratively, to arrive at another sine-cosine chain in one of the diagonal blocks of the squared Hamiltonian at each step. This subgroup of chains is labeled as $n$-times squarable sine-cosine chains [SSC($n$)], and the recursive procedure of squaring a SSC($n$) chain, applying an energy shift and hopping unit renormalisation until one reaches a uniform chain, is called the Matryoshka sequence \cite{Matryoshka}. 

This paper aims to expand the quantum state transfer protocol described in \cite{platero}, which focuses on the SSH chain with multiple domains, to the sine-cosine model in which an initial state completely localized at an end site of the chain evolves adiabatically to the other end through the mediation of topological edge states and intermediate domain wall states. 
Here, we propose to leverage the presence of edge states at different energy gaps of the SSC($n$) model to realize multiple transfer channels that can be used simultaneously. These HRTIs can be realized experimentally with photonic lattices \cite{wei2024realization,boross2019poor,perez2024transport}, where light injected in a waveguide has a probability of transitioning to their neighbouring guides depending on
the distance between them and on the refraction index contrast with the embedding crystal, and multiple transfer channels can be proven useful in recent (de)multiplexing strategies \cite{WU2024111997,guan2025chip}.
With the fragmentation of the SSC($2$) model into domain walls, we analyze the decay in transfer time, evaluate the transfer fidelity in two distinct ways and also explore the robustness of the topologically protected edge states against disorder in different regimes, as well as their transfer fidelity.

The rest of this paper is structured as follows. In Sec.~\ref{sec:section2}, we introduce the SSC($2$) chain as a 1D HRTI, paying special attention to its energy spectrum with two distinct gaps and the analytical expressions for the edge states contained within them. Sec.~\ref{sec:section3} focuses on the implementation of domain walls in the SSC($2$) chain and how this process leads to an effective model of the subspace spanned by the localized states. In Sec.~\ref{sec:section4}, we describe the quantum state transfer protocol with the adiabatic preparation of the states, derive the analytical expressions of the transfer time of the SSC($2$) and SSC($3$) chains with an arbitrary number of domains and map the relations of the transfer times both between gaps of the same root model and between root models. In Sec.~\ref{sec:section5}, we test the robustness of the SSC($2$) edge states against different disorder regimes and explore the impact of the disorder strength on the transfer fidelity of the model with varying number of domains. In Sec.~\ref{sec:section6}, we draw our conclusions.

\section{One dimensional high-root topological insulators} \label{sec:section2}
\begin{figure}[h!]
    \centering
    \includegraphics[width=\linewidth]{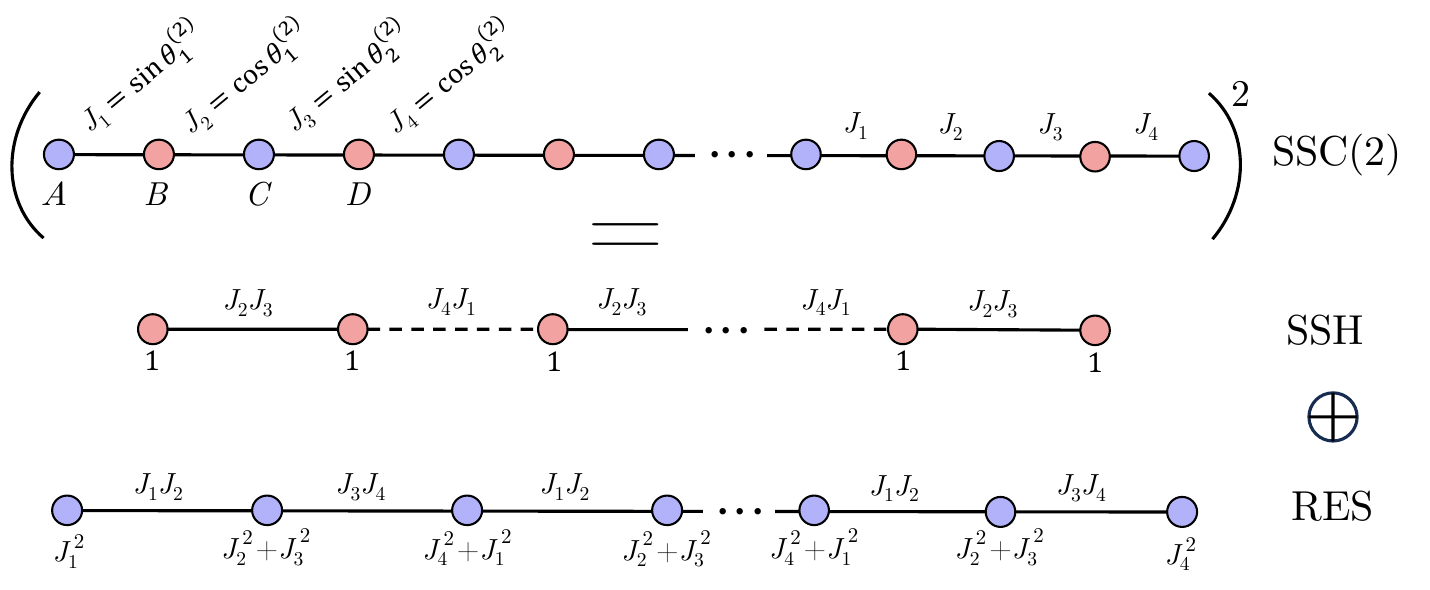}
    \caption{Squaring process of the SSC($2$) model into the parent SSH and residual chains with hopping parameters $\{J_1,J_2,J_3,J_4\}=\{\sin\theta_1^{(2)},\cos\theta_1^{(2)},\sin\theta_2^{(2)},\cos\theta_2^{(2)}\}$. The two different edge potentials in the residual chain result from the lower coordination number of the end sites.}
    \label{fig:squaringssc2}
\end{figure}

Fig.~\ref{fig:squaringssc2} depicts the squaring process of a SSC($2$) model with a length of $L=2l+1$ sites, where two chains will emerge, the residual (RES) chain with $l+ 1$ sites and the parent SSH chain with $N$ unit cells ($l=2N$ sites) and unitary on-site potentials.

\noindent The SSC($2$) Hamiltonian can be written as
\begin{align}
    \mathcal{H}=-\sum_{j=1}^N &\bigg\{J_1c^\dagger_{j,A}c_{j,B}+J_2c^\dagger_{j,B}c_{j,C} \nonumber\\
&+J_3c^\dagger_{j,C}c_{j,D}+J_4c^\dagger_{j,D}c_{j+1,A}+H.c.\bigg\},
\label{Hssc2}
\end{align}
where $c^\dagger_{j,\alpha}$ ($c_{j,\alpha}$) is the creation (annihilation) operator of a particle in site $\alpha\in\{A,B,C,D\}$ of the $j$-th unit cell. The parent SSH (residual) chain contains the sublattice composed of $B$ and $D$ ($A$ and $C$) sites of the SSC($2$) model, so by writing $\mathcal{H}$ in Eq.\,(\ref{Hssc2}) in matrix form, using the chiral basis $\{\{A\},\{C\},\{B\},\{D\}\}$, where $\{\alpha\}$ corresponds to the set of ordered $\alpha$ sites, one obtains a block anti-diagonal matrix of the form
\begin{equation}
    \mathcal{H}=\begin{pmatrix}
        \mathbb{0} & h\\
        h^\dagger& \mathbb{0}
    \end{pmatrix},
    \label{Hchiral}
\end{equation}
which squares to a block diagonal form,
\begin{equation}
    \mathcal{H}^2=\begin{pmatrix}
        hh^\dagger & \mathbb{0}\\
        \mathbb{0}& h^\dagger h
\end{pmatrix},
\end{equation}
with $hh^\dagger$ being the residual Hamiltonian and $h^\dagger h$ the parent SSH Hamiltonian up to a global energy shift of $J_1^2+J_2^2=J_3^2+J_4^2=1$. Given an arbitrary finite energy eigenstate $\ket{\psi_j}$ of the SSH Hamiltonian with energy $E_j$, we can apply $h$ to the left of both sides of its eigenvalue equation to obtain
\begin{equation}
h^\dagger h\ket{\psi_j}=E_j\ket{\psi_j}
    \Rightarrow hh^\dagger(h\ket{\psi_j})=E_j (h\ket{\psi_j}),
    \label{eq4}
\end{equation}
which results in $h\ket{\psi_j}$ being the (not-normalized) eigenstate of the residual Hamiltonian. The SSC($2$) eigenstates $|\psi_{\pm j}^{(2)}\rangle$ consist of a superposition of both residual and SSH eigenstates,
\begin{equation}
    |\psi_{\pm j}^{(2)}\rangle=\frac{1}{\sqrt{2}}\begin{pmatrix}E_j^{-1/2}h\ket{\psi_j}\\\pm\ket{\psi_j}\end{pmatrix},
    \label{chiraleq}
\end{equation}
which satisfy the SSC($2$) eigenvalue equation with energy $\pm\sqrt{E_j}$. Note that the Matryoshka sequence allows us to write the recurrence relations for the hopping parameters \cite{Matryoshka},
\begin{align}
&t^{(n-1)}\sin\theta^{(n-1)}_j=\cos\theta^{(n)}_{2j-1}\sin\theta^{(n)}_{2j},\label{rec1}\\
&t^{(n-1)}\cos\theta^{(n-1)}_j=\cos\theta^{(n)}_{2j}\sin\theta^{(n)}_{2j+1}\label{rec2},
\end{align}
where $n$ represents the order of the model, $j=1,...,2^{n-1}$ with $2^{n-1}+1\equiv 1$ and 
\begin{equation}
t^{(n-1)}=\sqrt{(\cos\theta^{(n)}_{2j-1}\sin\theta^{(n)}_{2j})^2+(\cos\theta^{(n)}_{2j}\sin\theta^{(n)}_{2j+1})^2}
\end{equation}
corresponds to the energy unit of the squared model, and also to derive the respective folding energies associated with the chiral symmetries (combination of particle-hole and time-reversal symmetries) that appear at each step of the squaring process,
\begin{align}
&\pm1,\nonumber\\
&\pm\sqrt{1\pm t^{(n-1)}},\nonumber\\
&\pm\sqrt{1\pm t^{(n-1)}\sqrt{1\pm t^{(n-2)}}}, \nonumber\\
&\vdots \nonumber\\
&\pm\sqrt{1\pm t^{(n-1)}\sqrt{1+\pm t^{(n-2)}\sqrt{\cdots\sqrt{1\pm t^{(1)}}}}}.\label{folding}
\end{align}

\begin{figure}[h!]
    \centering
    \includegraphics[width=\linewidth]{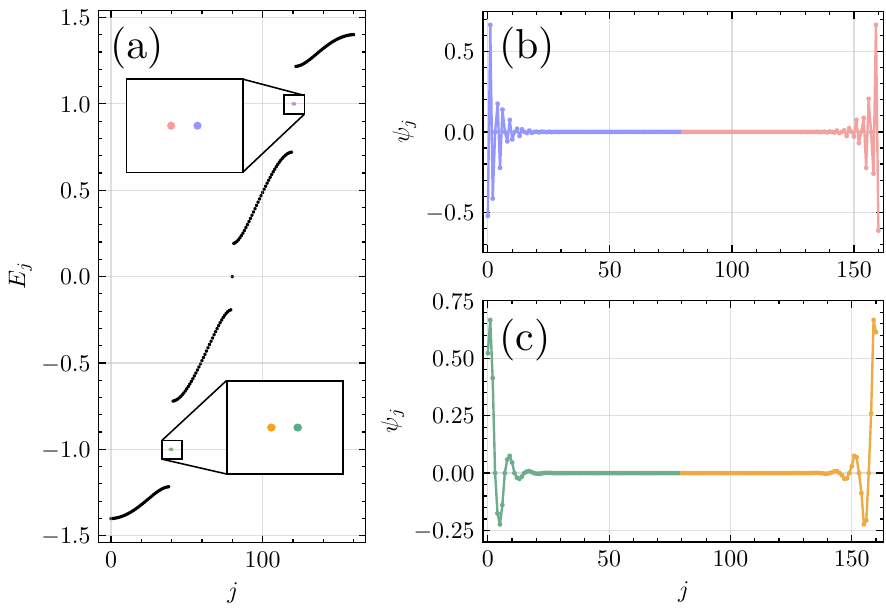}
    \caption{(a) Energy spectrum of the SSC($2$) chain with $\theta_1^{(2)}=0.286479\pi$, $\theta_2^{(2)}=0.127324\pi$ and 40 unit cells under OBC. (b) 
Probability amplitude at each site for the positive energy left edge state $|\mathcal{L}_+^{(2)}\rangle$ and right edge state $|\mathcal{R}_+^{(2)}\rangle$. (c) Probability amplitude at each site for the negative energy left edge state $|\mathcal{L}_-^{(2)}\rangle$ and right edge state $|\mathcal{R}_-^{(2)}\rangle$. The states in (b)-(c) and respective
energies in (a) are color-matched.}
    \label{fig:LRssc2}
\end{figure}

The edge states of any sine-cosine chain appear when a weak link is present at the boundaries \cite{Matryoshka}. After expressing the Hamiltonian in the chiral basis and applying the $h$ operator to the analytical expression of the parent SSH edge states in Eq.\,(\ref{eq4}) with intracell (intercell) hopping $J_2J_3$ ($J_4J_1$), the left and right edge states of the SSC($2$) chain are given by
\begin{align}
    |\mathcal{L}_\pm^{(2)}\rangle=&\frac{\mathcal{M}}{\sqrt{2}}\sum_{j=1}^N\Bigg\{\left(-\frac{J_4J_1}{J_2J_3}\right)^{-j}\times\nonumber\\&\times\left(J_1\ket{j,A}+J_2\ket{j,C}\mp\ket{j,B}\right)\Bigg\}
    \label{left}
\end{align}
and
\begin{align}
|\mathcal{R}_\pm^{(2)}\rangle=&\frac{\mathcal{M}}{\sqrt{2}}\sum_{j=1}^N\Bigg\{\left(-\frac{J_4J_1}{J_2J_3}\right)^{j-N-1}\times\nonumber\\&\times\left(J_3\ket{j,C}+J_4\ket{j+1,A}\mp\ket{j,D}\right)\Bigg\},
    \label{right}
\end{align}
respectively, where the sign in subscript indicates the energy $E=\pm1$ of the states and the normalization constant is given by $\mathcal{M}=\sqrt{(J_4J_1/J_2J_3)^2-1}$.
From the energy spectrum of the SSC(2) chain with $\theta_1^{(2)}=0.286479\pi$, $\theta_2^{(2)}=0.127324\pi$ and 40 unit cells depicted in Fig.~\ref{fig:LRssc2}(a), one can see the presence of two degenerate states within each energy gap, whose eigenstates are colormatched to the wavefunctions illustrated in Figs.~\ref{fig:LRssc2}(b-c). The absence of $D$ and $B$ components in the left and right edge states, respectively, comes
from the fact that in the parent SSH chain with open boundary conditions (OBC) these are the sublattice sites that do not
support them. 

Finally, as in the SSH model, edge states within the same energy gap and at opposite ends of the chain can hybridize if the conditions for the system size and hopping ratios are met, generating bonding and antibonding states, $(|\mathcal{L}_\pm^{(2)}\rangle+|\mathcal{R}_\pm^{(2)}\rangle)/\mathcal{M}_+$ and $(|\mathcal{L}_\pm^{(2)}\rangle-|\mathcal{R}_\pm^{(2)}\rangle)/\mathcal{M}_-$, respectively, whose normalization constants,
\begin{equation}
\mathcal{M}_\pm=\sqrt{2\pm\mathcal{M}^2J_2J_3\left(\frac{J_2J_3}{J_4J_1}\right)^{N+1}},
\end{equation}
take into account that the $|\mathcal{L}_+^{(2)}\rangle$ and $|\mathcal{R}_+^{(2)}\rangle$ edge states are not orthogonal as they overlap in the $A$ and $C$ sites.

\section{Domain walls}\label{sec:section3}
A domain wall is generated whenever two consecutive weak or strong links are present and cause a change in the dimerization pattern. The fact that these two chains are in distinct topological phases (one is trivial and the other non-trivial) results in the appearance of an in-gap localized state at that domain wall.
These domain wall states, upon hybridizing with the edge states, will serve as signal amplifiers and expedite the transfer process \cite{platero}. 

\subsection{Domain walls in the SSC($2$) chain}
\begin{figure}[h!]
    \centering
    \includegraphics[width=\linewidth]{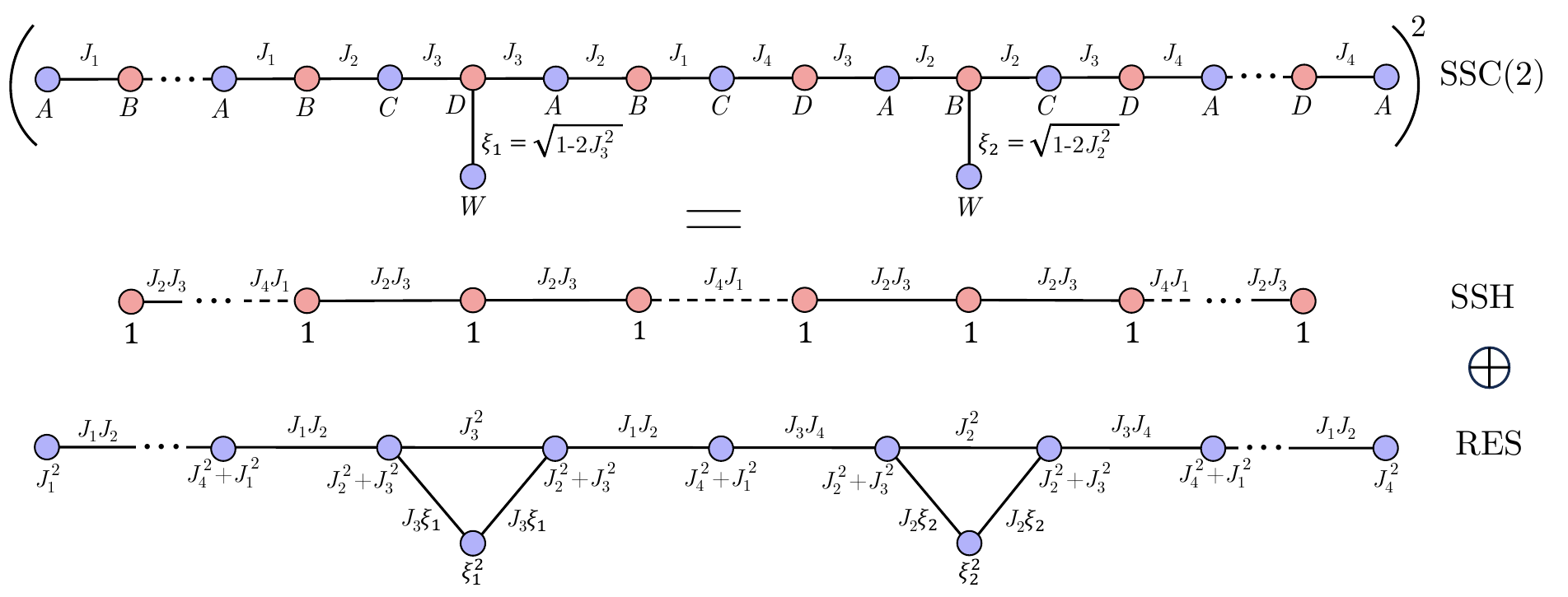}
    \caption{Squaring process of the multiple-domain SSC($2$) model into the parent SSH and residual chains. The two different edge potentials in the residual chain result from the lower coordination
number of the end sites.}
    \label{fig:squaringsscDdomains}
\end{figure}

Assuming that chiral symmetry is preserved, each domain wall state of the parent SSH will unfold into a pair of domain wall states of the SSC($2$) chain and one may use them as signal amplifiers as well.
Following the approach of \cite{marques2023kaleidoscopes}, we illustrate at the top of Fig.~\ref{fig:squaringsscDdomains} a SSC($2$) chain with an arbitrary number of domains, $d$, that leads to a parent SSH chain also fragmented by domain walls. 

\noindent One can see that a domain wall in the SSC($2$) model is formed whenever there is a twice repeating hopping terms and the middle site couples to a dangling site, and that the hopping sequence inverts from one domain to the next.
While the latter condition ensures that, upon squaring the SSC($2$) chain, each domain of the parent SSH chain possesses inversion symmetry, the first condition ensures that its domain wall sites have the same local potential as the other sites. Assuming the leftmost domain to have the index $w=1$, 
if the domain directly on the left of a domain wall has an odd (even) $w$, the corresponding dangling site is connected to a $B$ ($D$) site with the hopping $\xi_1=\sqrt{1-2J_3^2}$ ($\xi_2=\sqrt{1-2J_2^2}$). Note that upon squaring, the resulting parent model is a multiple-domain SSH chain with intracell (intercell) hopping $J_2J_3$ $(J_4J_1)$
and unitary onsite potentials, since the $B$ ($D$) domain wall sites have a local potential of $2J_3^2+\xi_1^2=1$ ($2J_2^2+\xi_2^2=1$). 
Consequently, the first angle is restricted by $\theta_1^{(2)}\in\left[\pi/4,3\pi/4\right]\cup\left[5\pi/4,7\pi/4\right]$ and the second one by $\theta_2^{(2)}\in \left[-\pi/4,\pi/4\right]\cup\left[3\pi/4,5\pi/4\right]$ in order to implement domain walls.
The Hamiltonian of an SSC($2$) chain with $d$ domains is written as
\begin{widetext}
\begin{align}
    \mathcal{H}_d=&-\sum_{w=1}^d \sum_{j=j_w^i}^{j_w^f}\Bigg\{\delta_{1,w\,\text{mod}\,2}\left[J_1c_{j,A}^\dagger c_{j,B}+J_2c_{j,B}^\dagger c_{j,C}+J_3c_{j,C}^\dagger c_{j,D}+(\delta_{j,j_w^f}-\delta_{w,d})(J_3c_{j,D}^\dagger c_{j+1,A}+\xi_1c_{j,D}^\dagger c_{j,W})\right.\nonumber\\
    &\left.+(1-\delta_{j,j_w^f}+\delta_{w,d})J_4c_{j,D}^\dagger c_{j+1,A}\right]+\delta_{0,w\,\text{mod}\, 2}
\left[J_2c_{j,A}^\dagger c_{j,B}+(1-\delta_{j,j_w^f})(J_1c_{j,B}^\dagger c_{j,C}+J_4c_{j,C}^\dagger c_{j,D}+J_3c_{j,D}^\dagger c_{j+1,A})\right.\nonumber\\
&\left.+(\delta_{j,j_w^f}-\delta_{w,d})(J_2c_{j,B}^\dagger c_{j,C}+\xi_2c_{j,B}^\dagger c_{j,W}+J_3c_{j,C}^\dagger c_{j,D}+J_4c_{j,D}^\dagger c_{j+1,A})+\delta_{w,d}J_1c_{j_w^f,B}^\dagger c_{j_w^f,C}\right]+h.c.\Bigg\}
\label{BigChungas}
\end{align}
\end{widetext}
where $\delta_{i,j}$ is the Kronecker delta, $j_w^i$ and $j_w^f$ correspond to the first and last unit cells of the domain $w$, respectively, and $W$ is the label of the dangling sites that will appear in the residual chain after squaring Eq.\,(\ref{BigChungas}).  
Since the domain wall states of the parent SSH chain are given by \cite{platero}  
\begin{align}
&|\mathcal{S}_w^{(1)}\rangle=\,\mathcal{M_S}\Bigg\{\sum_{j=j_w^i}^{j_w^f-1}\left(-\frac{J_4J_1}{J_2J_3}\right)^{j-j_w^f}|j,\Delta_w\rangle\nonumber\\
&+\sum_{j=j_{w+1}^i}^{j_{w+1}^f}\left(-\frac{J_4J_1}{J_2J_3}\right)^{j_w^f-j}|j,\Delta_w\rangle +|j_w^f,\Delta_w\rangle\Bigg\},
    \label{domainstatessh}
\end{align}

\begin{figure}[h!]
\centering
\includegraphics[width=\linewidth]{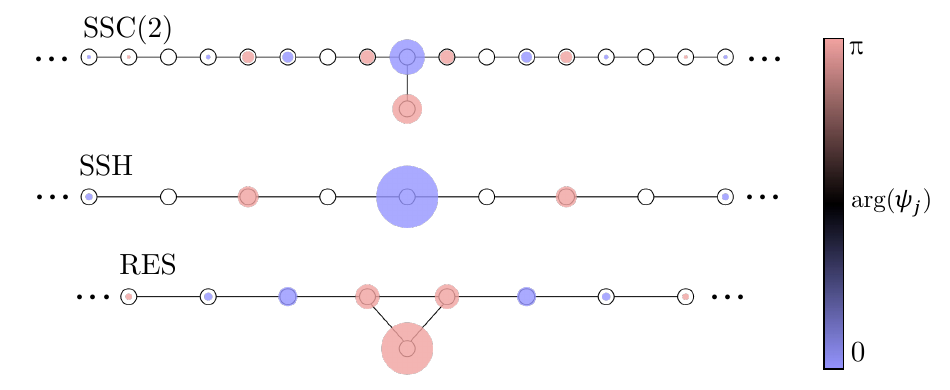}
\caption{First domain wall state of the SSC(2) model with length $L=112$, $d=2$ domains with $14$ unit cells each, $\theta_1^{(2)}=0.286479\pi$, $\theta_2^{(2)}=0.127324\pi$ and the corresponding normalized weights in the parent SSH and residual chains. The size and color of the disks are proportional to their amplitude and phase, respectively.}
\label{ssc2_S1}
\end{figure}

\noindent with $w=1,2,...,d-1$, $\Delta_w=D$ ($\Delta_w=B$) for odd (even) $w$ and normalization factors given by $\mathcal{M_S}=\sqrt{(J_4^2-J_2^2)/(J_4^2+J_2^2-2J_4^2J_2^2)}$,
the domain wall states of the SSC($2$) model can be obtained through Eq.\,(\ref{chiraleq}), 
\begin{align}
&h|\mathcal{S}_w^{(1)}\rangle=-\mathcal{M_S}\delta_{1,w\,\text{mod}\,2}\Bigg\{J_3|j_w^f+1,A\rangle+\xi_1|j_w^f,W\rangle\nonumber\\
    &+\sum_{j=j_w^i}^{j_w^f-1}\left(-\frac{J_4J_1}{J_2J_3}\right)^{j-j_w^f}(J_3|j,C\rangle+J_4|j+1,A\rangle)\nonumber\\
    &+\sum_{j=j_{w+1}^i}^{j_{w+1}^f}\left(-\frac{J_4J_1}{J_2J_3}\right)^{j_w^f-j}(J_4|j,C\rangle+J_3|j+1,A\rangle)\Bigg\}\nonumber\\
    &-\mathcal{M_S}\delta_{0,w\,\text{mod}\,2}\Bigg\{J_2|j_w^f,C\rangle+\xi_2|j_w^f,W\rangle\nonumber\\
    &+\sum_{j=j_w^i}^{j_w^f-1}\left(-\frac{J_4J_1}{J_2J_3}\right)^{j-j_w^f}(J_2|j,A\rangle+J_1|j,C\rangle)\nonumber\\
    &+\sum_{j=j_{w+1}^i}^{j_{w+1}^f}\left(-\frac{J_4J_1}{J_2J_3}\right)^{j_w^f-j}(J_1|j,A\rangle+J_2|j,C\rangle)\Bigg\}.
    \label{hS}
\end{align}

Fig.~\ref{ssc2_S1} illustrates the first domain wall state of the SSC($2$) model with multiple domains as well as the corresponding individual contributions from the parent SSH and residual chains.

\subsection{Effective model of the SSC($2$) chain}

\begin{figure}[h!]
    \includegraphics[width=0.9\linewidth]{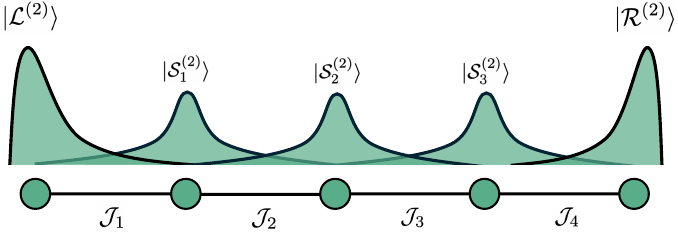}    
    \caption{Effective model of the SSC($2$) chain with four domains.}
    \label{fig:effective}
\end{figure}

In-gap states, well separated from bulk states, form an effective decoupled subspace up to leading order in perturbation theory. This subspace translates into an effective model that is capable of describing the dynamical properties of the topological boundary states. The effective model of the SSC($2$) chain is depicted in Fig.\,\ref{fig:effective}, where the sites represent the edge and domain wall states and the effective hoppings are given by $\mathcal{J}_w=\langle \mathcal{S}_{w-1}^{(2)}|\mathcal{H}_d|\mathcal{S}^{(2)}_w\rangle$, with $| \mathcal{S}_{0}^{(2)}\rangle\equiv|\mathcal{L}^{(2)}\rangle$ and $| \mathcal{S}_{d}^{(2)}\rangle\equiv|\mathcal{R}^{(2)}\rangle$. Note that inversion symmetry between each pair of consecutive domains and the form of the domain wall states impose $\mathcal{J}_1=\mathcal{J}_d$ and $\mathcal{J}_{w}=\mathcal{J}_{w+1}$ for $1<w<d-2$, respectively.

\begin{figure}[h!]
\includegraphics[width=\linewidth]{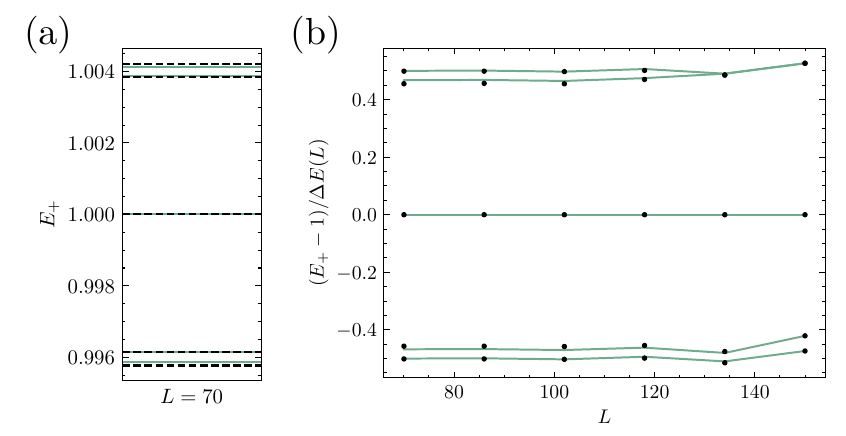}    
\caption{(a) Energy spectrum of the positive energy in-gap states of a four-domain SSC($2$) model with $\theta_1^{(2)}=0.286479\pi$, $\theta_2^{(2)}=0.127324\pi$ and $L=70$ with $4$ unit cells per domain. (b) Renormalized energy spectrum of the positive in-gap states as a function of the size of the four-domain SSC($2$) chain, where $\Delta E(L)$ is the bandwidth of the subspace of positive energy in-gap states for each $L$. The results in black are obtained from the exact diagonalization of the full Hamiltonian given in Eq.\,(\ref{BigChungas}) and the results in green from the energies given by the perturbative expansion of Eq.\,(\ref{Eeff}).}
\label{fig:perturbation}
\end{figure}

Recalling the structure of a general eigenstate of the SSC($2$) model given in Eq.\,(\ref{chiraleq}) with $E_j=\pm1$ and the absence of weights in $D$ sites in the left edge state of [Eq.\,(\ref{left})], one can obtain the first effective hopping by writing its overlap in the chiral basis,
\begin{align}
\mathcal{J}_1&=\frac{1}{\sqrt{2}}\begin{pmatrix}
\langle \mathcal{L}_{A/C}^{(2)}| & \langle \mathcal{L}_{B/D}^{(2)}|
\end{pmatrix}\begin{pmatrix}
0 & h\\
h^\dagger & 0
\end{pmatrix}\begin{pmatrix}
h| \mathcal{S}_{1}^{(1)}\rangle \\
|\mathcal{S}_{1}^{(1)}\rangle
\end{pmatrix}\nonumber\\
&\approx\frac{\mathcal{MM}_S}{2}\bigg\{
\sum_{j=j_2^i}^{j_2^f}\left(-\frac{J_4J_1}{J_2J_3}\right)^{j_1^f-2j}\left(J_4J_2-\frac{J_2J_3^2}{J_4}\right.\nonumber\\
&+\left.J_4J_1-\frac{(J_2J_3)^2}{J_4J_1}\right)-J_1J_3\left(-\frac{J_4J_1}{J_2J_3}\right)^{-j_1^f-1}\bigg\},
\end{align}
and the same procedure holds for the inner effective hoppings,
\begin{align}
\mathcal{J}_w&=\frac{1}{2}\begin{pmatrix}
\langle \mathcal{S}_{w-1}^{(2)}|h^\dagger & \langle \mathcal{S}_{w-1}^{(2)}|
\end{pmatrix}\begin{pmatrix}
0 & h\\
h^\dagger & 0
\end{pmatrix}\begin{pmatrix}
h| \mathcal{S}_{w}^{(1)}\rangle \\
|\mathcal{S}_{w}^{(1)}\rangle
\end{pmatrix}\nonumber\\
&\approx\mathcal{M}_S^2\frac{J_2J_4}{J_1J_3} (J_2-J_1)\left(-\frac{J_4J_1}{J_2J_3}\right)^{j_{w-1}^f-j_{w}^f},
\end{align}
where $1<w<d-1$. At the thermodynamic limit, all positive (negative) in-gap states of the SSC($2$) chain have energies $E=1$ ($E=-1$) regardless of the number of domains. As such, one can treat the effective Hamiltonian 
\begin{equation}
\tilde{\mathcal{H}}_{d,\pm}=\pm\mathbb{1}_d-\sum_{w=1}^d\mathcal{J}_w |\mathcal{S}_{w-1}^{(2)}\rangle\langle\mathcal{S}_{w}^{(2)}|+H.c.
\label{Heff}
\end{equation}
with $\mathbb{1}_d$ being the identity matrix of size $d$, as a perturbation to these states and thus the perturbed energies of the in-gap states are given by
\begin{equation}
E_\pm=\pm(1+\tilde{E}),
\label{Eeff}
\end{equation}
where the sign in subscript refers to wether we are considering the positive or negative energy in-gap states and $\tilde{E}$ are the eigenenergies of Eq.\,(\ref{Heff}). Fig.\,\ref{fig:perturbation} presents the energy spectrum of the positive energy in-gap states of a four-domain SSC($2$) model with $L=70$ and the renormalized energy spectrum of these states for increasing system size where $\Delta E(L)=\text{max}(E_+)-\text{min}(E_+)$ for each $L$.
 One can see that the results obtained from exact diagonalization of the full Hamiltonian given in Eq.\,(\ref{BigChungas}) (black) and the energies given by the perturbative expansion of Eq.\,(\ref{Eeff}) (green) are in good agreement.

\begin{figure*}
    \includegraphics[width=\linewidth]{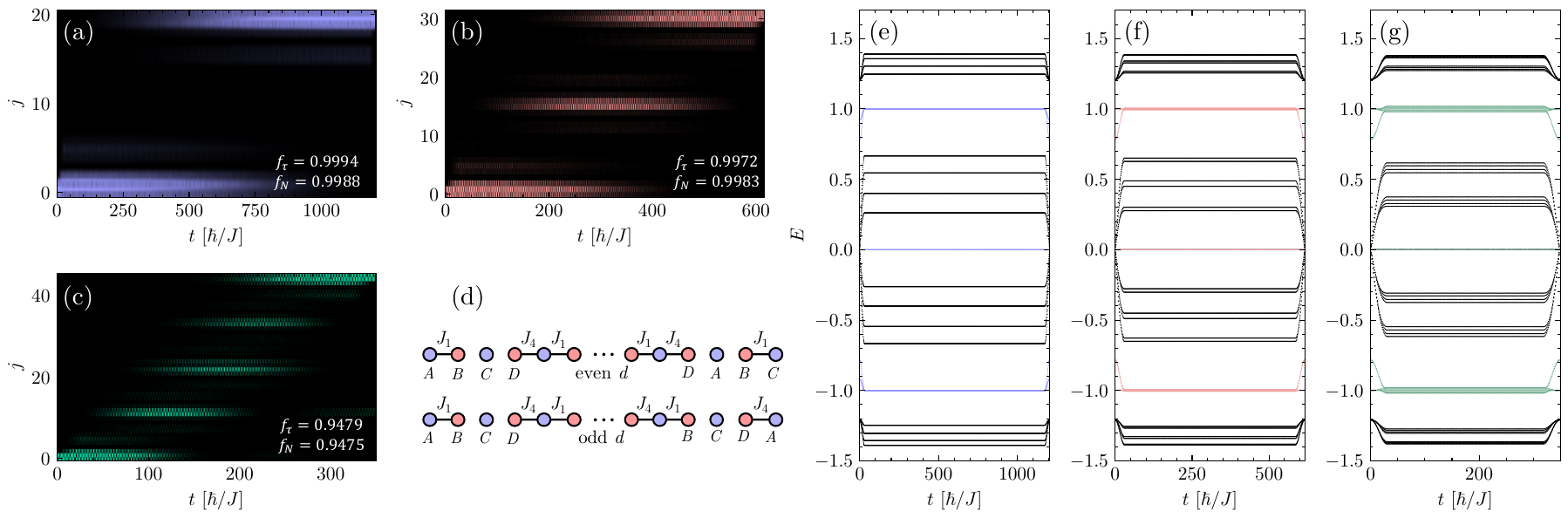}    
    \caption{Probability per site index $|\psi_j|^2$ over time and fidelities of the SSC($2$) chains with $\theta_1^{(2)}=0.286479\pi$ and $\theta_2^{(2)}=0.127324\pi$ for the case of one domain with (a) 21 sites, (b) two domains with 32 sites, and (c) four domains with 46 sites. (d) Representation of the SSC($2$) chain at the beginning and end of the transfer process for even and odd number of domains. Energy spectrum of the model with (e) one domain, (f) two domains and (g) four domains where the energies of the edge states are color-matched with the respective model.}
    \label{fig:transferall}
\end{figure*}

\section{Quantum transfer} \label{sec:section4}
\subsection{Transfer protocol}
To extend the adiabatic state preparation and transfer realized in \cite{platero} to the SSC($2$) model, we apply the following time-dependent adiabatic factor,
\begin{equation}
\sigma(t)=\begin{cases}
    |\sin(\Omega t)|\hspace{47pt}\text{for }0\leq t<t_\text{prep},\\
    1\hspace{80pt}\text{for }t_\text{prep}\leq t<t_\text{tr}-t_\text{prep},\\
    |\sin[\Omega(t-t_\text{prep})]|\hspace{10pt}\text{for }t_\text{tr}-t_\text{prep}\leq t<t_\text{tr},
\end{cases}
\label{sigma}
\end{equation}
where $\Omega=\pi/2t_\text{prep}$ is the preparation frequency, to each of the hoppings $J_2$ and $J_3$, since $J_2J_3\sigma^2$ translates into the intracell parameter of the parent SSH chain. 
The initial state is a state completely localized in the leftmost site of the chain and, during the preparation period, the particle oscillates between the first two sites since they are connected to the rest of the chain by an intracell hopping that is still being tuned up [Fig.~\ref{fig:transferall}(b)], forming the Rabi oscillation pattern seen in Fig.~\ref{fig:transferall}(a).
During the preparation period, the initial state evolves into a superposition of $|\mathcal{L}_+^{(2)}\rangle$ and $|\mathcal{L}_-^{(2)}\rangle$ due to their high amplitude in the leftmost sites and, during the transfer period, each one evolves into respective counterpart state at the right edge $|\mathcal{R}_\pm^{(2)}\rangle$. After the system state becomes the superposition of $|\mathcal{R}_+^{(2)}\rangle$ and $|\mathcal{R}_-^{(2)}\rangle$, the model undergoes another adiabatic period to constrain the state in the two rightmost sites of the chain, allowing the measurement to take place at the last dimer \cite{wei2024realization}. 
From now on, we will express time in units of $\hbar/J$, where $J$ is the energy unit, such that the SSC($n$) hoppings range from 0 to 1, and we considered $\hbar\equiv1$.

We optimized the parameters that returned the best transfer fidelity results, setting the optimal preparation time $t_\text{prep}=30\hspace{2pt}\hbar/J$ and applying the time evolution operator $e^{-i\mathcal{H}(t)t/\hbar}$ to the initial state in time steps of $dt=0.1\, \hbar/J$ on all simulations. We evaluate the fidelity of the transfer in two distinct ways: one is the internal fidelity $f_\tau=|\langle \mathcal{R}_\pm^{(2)}|\psi(t_\text{prep}+\tau)\rangle|$, where $\tau$ is the instant at which $e^{-i\mathcal{H}\tau/\hbar}|\mathcal{L}_\pm^{(2)}\rangle=|\mathcal{R}_\pm^{(2)}\rangle$, and the other is the fidelity at the last dimer $|N\rangle=|L-1\rangle+|L\rangle$, $f_N=|\langle N|\psi(t_\text{tr})\rangle|$. In light of this, the transition between internal and final adiabatic period occurs when $f_\tau$ reaches its maximum value.

Fig.~\ref{fig:transferall}(a-c) illustrates the transfer process of a SSC($2$) chain for the cases of $d=1,2,4$ domains with $L=21,32,46$ sites, respectively. Although the transfer time decreases exponentially after incorporating domain walls, a fragment of the state that is being transferred becomes confined in each domain wall, oscillating between the domain wall site and the attached dangling site, which decreases the transfer fidelity as observed in the first domain wall at the end of the transfer of Fig.~\ref{fig:transferall}(c). 
At the beginning and end of the transfer process, the SSC($2$) chain to fall into trimers, except in the first and last unit cells as observed in Fig.~\ref{fig:transferall}(d). This limit forces the left edge state to always have an energy of $J_1$ and the right edge state to have an energy of $J_4$ for an odd number of domains (Fig.~\ref{fig:transferall}(e)) and $J_1$ for an even number of domains due to reflection symmetry upon each domain wall (Figs.~\ref{fig:transferall}(f)-(g)) and the domain wall states to have an energy of $E=1$.

\begin{figure}[h!]
    \centering
    \includegraphics[width=\linewidth]{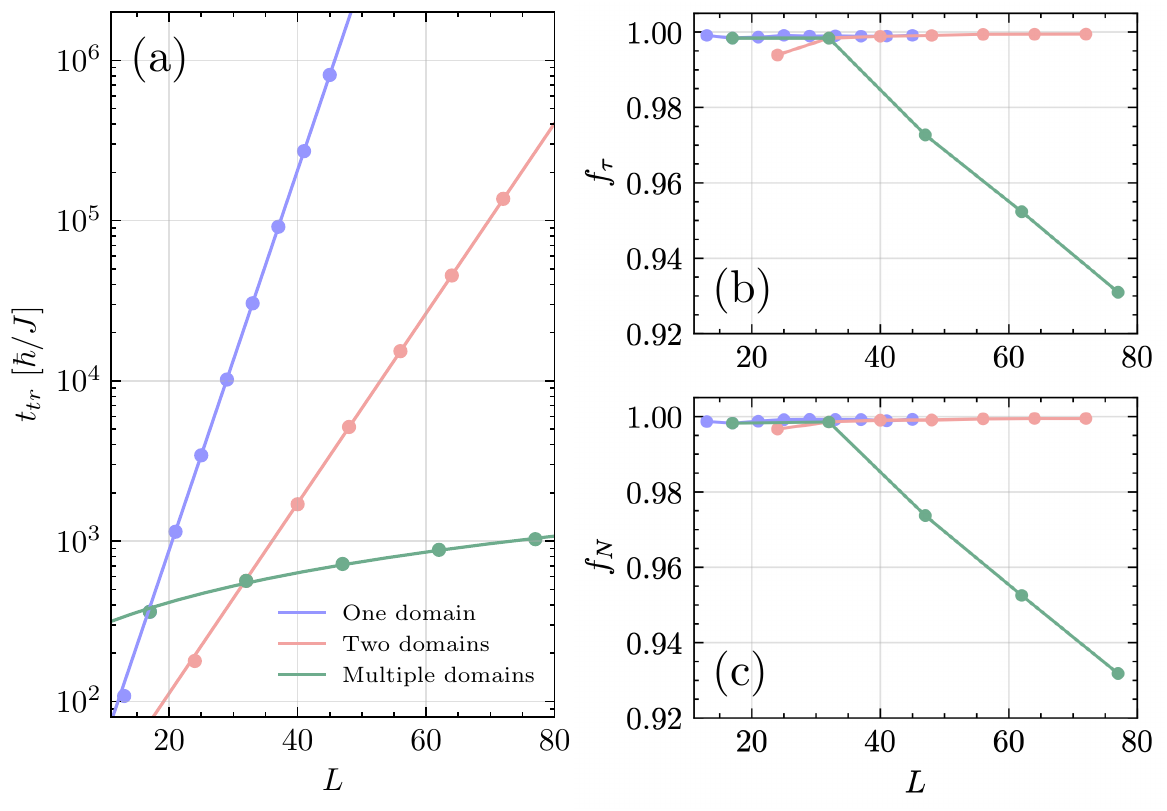}    
    \caption{(a) Transfer time, (b) internal fidelity and (c) last dimer fidelity of the SSC($2$) model as a function chain length for the cases of one, two and multiple domains. The results were obtained using $\theta_1^{(2)}=0.286479\pi$ and $\theta_2^{(2)}=0.127324\pi$, where the dots correspond to the numerical results of the adiabatic transfer protocol, the purple and pink lines to two preparation periods plus the theoretical predictions of Eqs.\,(\ref{analyticaltau1}-\ref{analyticaltau2}), respectively, and the green line to the linear fit $11L+195$.}
    \label{transfer}
\end{figure}

For a single-domain SSC($2$) chain, the probability of the left edge state in Eq.\,(\ref{left}) evolving into its equivalent right edge state given by Eq.\,(\ref{right}) is maximum when $\tau=\pi\hbar/\Delta E_\pm$, with $\Delta E_\pm$ being the energy gap between their symmetric and anti-symmetric combinations (which are the true eigenstates) given by
\begin{align}
\Delta E_\pm&=\frac{1}{|\mathcal{M}_+|^2}\left(\langle\mathcal{L}^{(2)}_+|+\langle\mathcal{R}^{(2)}_+|\right)\mathcal{H}\left(|\mathcal{L}^{(2)}_+\rangle+|\mathcal{R}^{(2)}_+\rangle\right)\nonumber\\
&-\frac{1}{|\mathcal{M}_-|^2}\left(\langle\mathcal{L}^{(2)}_+|-\langle\mathcal{R}^{(2)}_+|\right)\mathcal{H}\left(|\mathcal{L}^{(2)}_+\rangle-|\mathcal{R}^{(2)}_+\rangle\right)\nonumber\\
&=\mathcal{M}^2J_2J_3\left(\frac{J_2J_3}{J_4J_1}\right)^{(L+3)/4}
\end{align}
thus the internal transfer time of a one-domain SSC($2$) is given by
\begin{equation}
    \tau^{(2)}_{d=1}(L)=\frac{\pi\hbar}{\mathcal{M}^2J_2J_3}\left(\frac{J_4J_1}{J_2J_3}\right)^{(L+3)/4}.
    \label{analyticaltau1}
\end{equation}
As expected, the transfer time increases exponentially with the length of the chain. 
Note that the internal transfer time of a left edge state in the SSC($2$) model given in Eq.\,(\ref{analyticaltau1}) is twice the one of a SSH model with intracell (intercell) hopping $J_2J_3$ $(J_4J_1)$ obtained in \cite{platero} for the same number of unit cells, a consequence of the energy relation within the SSC($n$) family \cite{Matryoshka} that is fully explained in Sec.~\ref{subsectiontransfer}.
With this in mind, we can write the expression for the internal transfer time of the two-domain SSC($2$) chain with length $L$ as
\begin{equation}
\tau^{(2)}_{d=2}(L)=\frac{\pi\hbar\sqrt{2\mathcal{M}^2+4}}{\mathcal{M}^2J_2J_3}\left(\frac{J_4J_1}{J_2J_3}\right)^{L/8},
\label{analyticaltau2}
\end{equation}
based on the internal transfer time of its two-domain parent SSH chain with intracell (intercell) hopping $J_2J_3$ $(J_4J_1)$ and length $L'=L/2-1$ obtained in \cite{platero},
\begin{equation}
\tau^{(1)}_{d=2}(L')=\frac{\pi\hbar\sqrt{(J_4J_1)^2+(J_2J_3)^2}}{\sqrt{2}((J_4J_1)^2-(J_2J_3)^2)}\left(\frac{J_4J_1}{J_2J_3}\right)^{(L'+1)/4}.
\end{equation}
We can conclude that the transfer time for a two-domain SSC($2$) chain also increases exponentially with length but slower than a one-domain chain [see Eq.\,(\ref{analyticaltau1})], since the exponential coefficient $L/8$ is smaller than $(L+3)/4$, meaning that the domain wall provides a faster transfer process for the left edge state. This means that if we fix the inner length of each domain and increase the number of domains in the chain, the internal transfer time will become linearly dependent on the size of the chain, instead of exponentially dependent. 
Fig.~\ref{transfer}(a) illustrates the transfer time of the SSC($2$) model as a function of chain length for the cases of one, two and multiple domains, where the dots correspond to the numerical results of the adiabatic transfer protocol, the purple and pink lines to two preparation periods plus the theoretical predictions of Eqs.\,(\ref{analyticaltau1}-\ref{analyticaltau2}), respectively, and the green line to the linear fit $11L+195$. Despite the transfer being exponentially faster for the case of multiple domains, Figs.~\ref{transfer}(b-c) show a rapid decay in transfer fidelity due to the partial state leakage to the domain wall sites along the transfer process. 

\subsection{Transfer time mapping}\label{subsectiontransfer}
The presence of an edge state in each of the $2^{n-1}$ band gaps of the SSC($n$) model allows for a state transfer across $2^{n-1}$ channels simultaneously. One can express the transfer time between the states lying within the $m$-th gap ($m\in\mathbb{Z}^+\leq2^{n-2}$) of the SSC($n$) model as 
\begin{equation}
\tau_m^{(n)}=\frac{\pi\hbar}{E_{m,+}^{(n)}-E_{m,-}^{(n)}},
\label{taumn}
\end{equation}
where $E_{m,+}^{(n)}$ ($E_{m,-}^{(n)}$) corresponds to the antibonding (bonding) energy. The folding levels defined in Eq.\,(\ref{folding})  and the recurrence relations of Eqs.\,(\ref{rec1}) and (\ref{rec2}) enable us to write 
\begin{equation}
    E_{m,\pm}^{(n)}=\sqrt{1+\Gamma_{m,n-1}t^{(n-1)}E_{\ceil{m/2},\pm}^{(n-1)}}
    \label{recrel}
\end{equation}
where $\ceil{x}$ is the ceiling function and the $\Gamma$ function is defined as
\begin{equation}
\Gamma_{\alpha,\beta}:=\prod_{\lambda=\beta}^{n-1}\bigg\{2\Theta(\alpha-2^{\lambda-2}-1)-1\bigg\}
\end{equation}
in which the product ensures the correct sign change under reflection symmetry of the folding levels and $\Theta(x)$ is the Heaviside step function. As such, Eq.\,(\ref{recrel}) ends with the bonding and antibonding energies of the parent SSH chain given by
\begin{equation}
E_{1,\pm}^{(1)}=\pm\frac{\cos2\theta_1^{(1)}}{\sin\theta_1^{(1)}}\left(\tan\theta_1^{(1)}\right)^{L/2+1},
\end{equation}
where the angle $\theta_1^{(1)}$ can be extracted from the coupled expressions and the length of the parent SSH chain, $L$, is related to the length of the SSC($n$) chain by $L^{(n)}=2^{n-1}(L+1)+1$.
Since the bonding and antibonding energies are close to the folding levels [see Eq.\,(\ref{folding})], one can expand Eq.\,(\ref{recrel}) around the corresponding folding energy, yielding
\begin{equation}
E_{m,\pm}^{(n)}\approx\epsilon_m^{(n)}+\frac{t^{(n-1)}}{2\epsilon_m^{(n)}}\left(E_{\ceil{m/2},\pm}^{(n-1)}-\epsilon_m^{(n-1)}\right),
\label{approxE}
\end{equation}
where $\epsilon_m^{(j)}=\sqrt{1+\Gamma_{m,j-1}t^{(j-1)}\epsilon_{\ceil{m/2}}^{(j-1)}}$ and $\epsilon_1^{(1)}=0$, leading to the recurrence relation
\begin{equation}
    \tau_m^{(n)}=\frac{2\epsilon_m^{(n)}}{t^{(n-1)}}\tau_{\ceil{m/2}}^{(n-1)},
    \label{recdiffroot}
\end{equation}
where we have used Eq.\,(\ref{taumn}). It is also possible to establish the relation between the transfer times of different band gaps defined by $m$ and $m'$ of the same SSC($n$) model as
\begin{equation}
\tau_m^{(n)}=\tau_{m'}^{(n)}\prod_{j=3}^n\frac{\epsilon_{\ceil{m/2^{j-3}}}^{(j)}}{\epsilon_{\ceil{m'/2^{j-3}}}^{(j)}},\label{recsameroot}
\end{equation}
for $n\geq 3$ given that the SSC($3$) model is the first root model containing more than one positive energy gap, completing the transfer time mapping between root models and different band gaps of the same model. Note that this approach is valid for an arbitrary number of domains since it only depends on the values of the respective energy gaps.

\subsection{Quantum state transfer in the SSC($3$) model}

One can use the recurrence relations defined in Eqs.\,(\ref{rec1}) and (\ref{rec2}) in order to obtain the SSC($3$) model whose Hamiltonian satisfies
\begin{equation}
H_k^{(3)}=\pm\sqrt{H_k^{(2)}t^{(2)}+\mathbb{1}}.
\end{equation}

\begin{figure}[h!]
    \includegraphics[width=\linewidth]{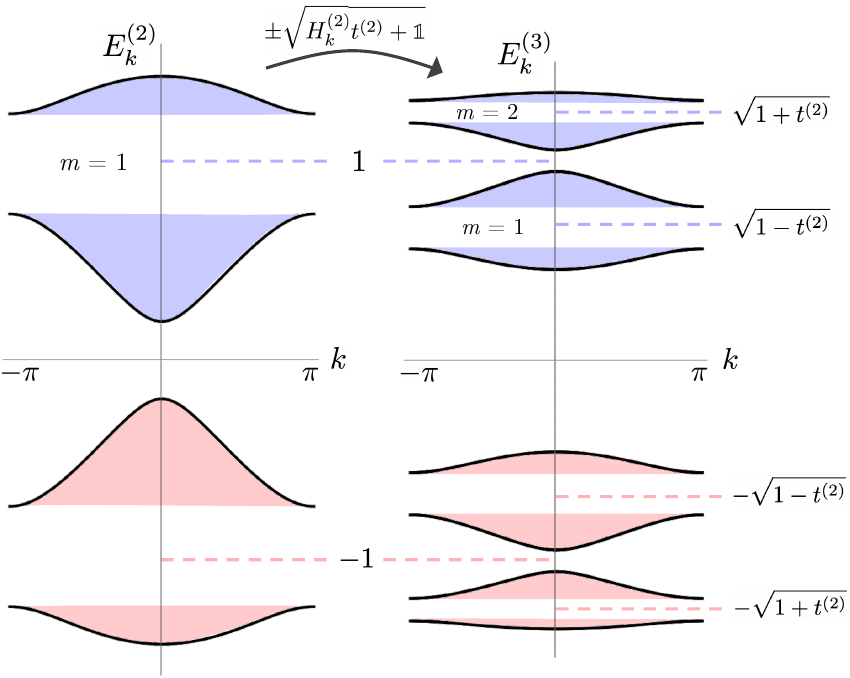}    
    \caption{Inverse Matryoshka sequence for arriving at the energy spectrum of the SSC($3$) model (at the right) from the SSC($2$) model (at the left).}
    \label{ssc3bands}
\end{figure}

Since the SSC($2$) chain contains three folding levels ($0$ and $\pm1$) associated to chiral symmetry and only one unique band gap in the positive energy region, the SSC($3$) will possess six folding energies ($\pm1$ and $\pm\sqrt{1\pm t^{(2)}}$) which unfold two unique band gaps in the positive energy region separated by a gap containing the folding level $+1$, as depicted in Fig.\,\ref{ssc3bands}. For a set of angles that satisfies the existence of a weak link in the hopping $\sin\theta_4^{(3)}$, each of these two unique gaps will contain a pair of energy symmetric edge states that can be used in quantum state transfer protocols and, since the transfer process depends on the energy gap in which the initial state is contained, there will be two possible transfer times. 

\begin{figure}[h!]
    \includegraphics[width=\linewidth]{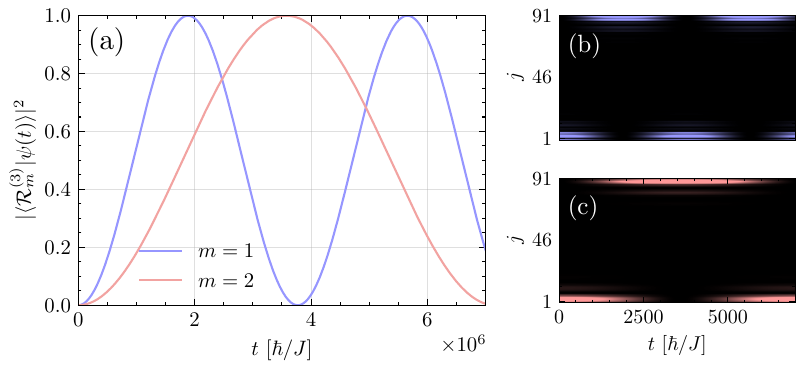}    
    \caption{(a) Probability evolution of the initial state transferring to the right edge state of the same gap $m=1,2$ in the SSC($3$) model. Probability per site index $|\psi_j|^2$ over time of the SSC($3$) chain with initial state in gap (b) $m=1$ and (c) $m=2$. The results were obtained for a chain with $\{\theta_1^{(3)},\theta_2^{(3)},\theta_3^{(3)},\theta_4^{(3)}\}=\{0.566852, 0.553034, 0.426003, 0.244332\}$, $t^{(2)}=0.565685$ and 91 sites.}
    \label{fig:ssc3transfer}
\end{figure}

Without performing the state preparation protocol and considering the angles $\{\theta_1^{(3)},\theta_2^{(3)},\theta_3^{(3)},\theta_4^{(3)}\}=\{0.566852, 0.553034, 0.426003, 0.244332\}$ that satisfy the recurrence relations of Eqs.\,(\ref{rec1}) and (\ref{rec2}) with $\theta_1^{(2)}=0.286479\pi$, $\theta_2^{(2)}=0.127324\pi$ and $t^{(2)}=0.565685$, Fig.\,\ref{fig:ssc3transfer}(a) shows the probability of the initial state transferring to the right edge state of the same gap $m=1,2$ over time, where we can see the existence of two complete transfers between the boundaries as observed in Fig.\,\ref{fig:ssc3transfer}(b), as well as two different transfer times. 
From Eq.\,(\ref{recsameroot}), we arrive at the relation between the transfer times of this model as 
\begin{equation}
\tau_1^{(3)}=\tau_2^{(3)}\sqrt{\frac{1-t^{(2)}}{1+t^{(2)}}},
\end{equation}
which is in agreement with the results $\tau_1^{(3)}=1.88407\times10^6\,\hbar/J$ and $\tau_2^{(3)}=3.57684\times10^6\,\hbar/J$ shown in Fig.\,\ref{fig:ssc3transfer}(a). On the other hand, from Eq.\,(\ref{recdiffroot}) we can relate the transfer time of the only positive energy gap of the parent SSC($2$) model with the transfer time of the first gap ($m=1$) of the SSC($3$) chain through
\begin{equation}
\tau_1^{(2)}=\frac{t^{(2)}}{2 \sqrt{1-t^{(2)}}}\tau_1^{(3)}\approx 8.08\times10^5\,\hbar/J
\end{equation}
which is very close to the analytical result of $\tau_1^{(2)}=8.09\times10^5\,\hbar/J$ from Eq.\,(\ref{analyticaltau1}) with $L=45$, thus confirming the theoretical expressions for the transfer time mapping.

\break

\begin{figure}[h!]
    \includegraphics[width=\linewidth]{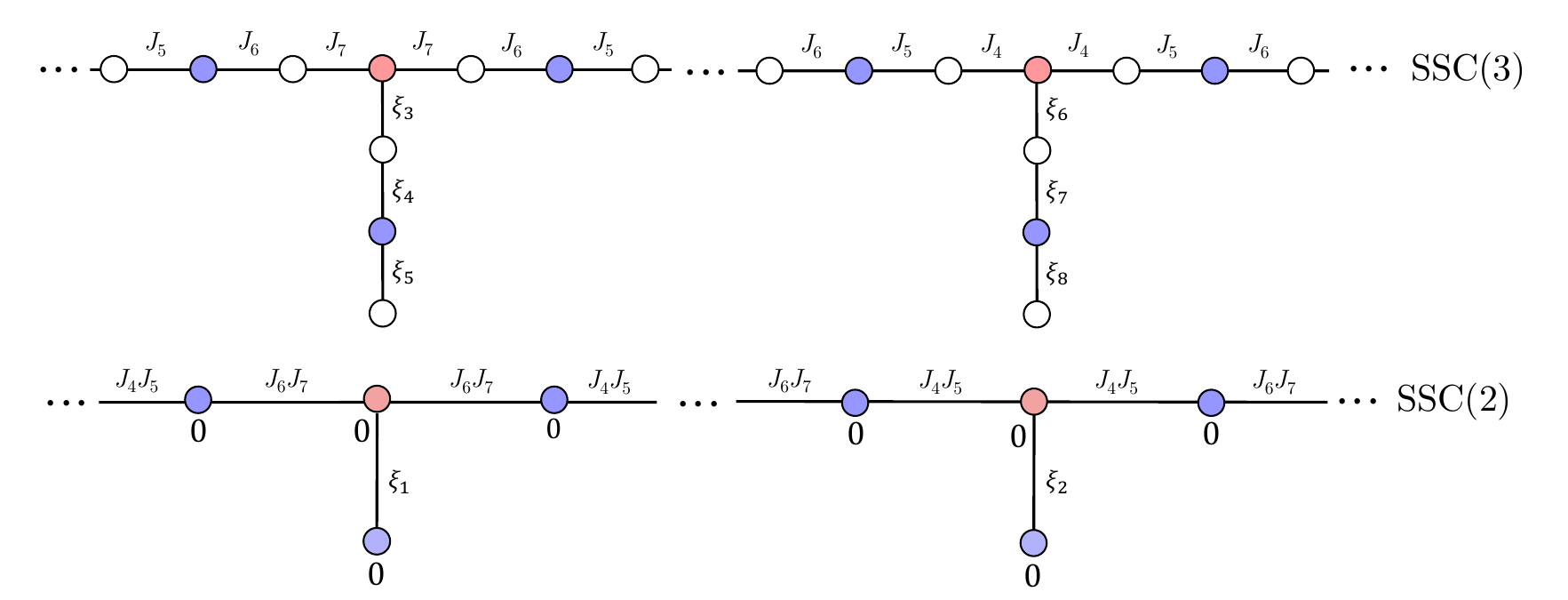}    
    \caption{Matryoshka sequence of the multiple-domain SSC($3$) model with the hopping parameters $J_{2j-1}=\sin\theta_j^{(3)}$ and $J_{2j}=\cos\theta_j^{(3)}$, where $j=1,2,3,4$.}
    \label{fig:ssc3domains}
\end{figure}

Following a similar approach to that in Sec.\,\ref{sec:section3}, it is possible to implement domains walls in the SSC($3$) model, as shown in  Fig.\,\ref{fig:ssc3domains}. 
The dangling sites are now replaced by dangling chains, which not only preserves inversion-symmetry between adjacent domains, but also the unitary on-site potentials in all sites (including the dangling sites) of the parent SSC($2$) model.
Relating the multiple-domain SSC($3$) model with its parent SSC($2$) model, one gets $\xi_1=\sqrt{(t^{(2)})^2-2(J_6J_7)^2}$, $\xi_2=\sqrt{(t^{(2)})^2-2(J_4J_4)^2}$ and the hoppings of the first and second dangling chain must be
\begin{equation}
\xi_3=\sqrt{1-2J_7^2},\,\,\,\,\,\, \xi_4=\xi_1/\xi_3,\,\,\,\,\,\, \xi_5=\sqrt{1-\xi_4^2}
\end{equation}
and 
\begin{equation}
\xi_6=\sqrt{1-2J_4^2},\,\,\,\,\,\, \xi_7=\xi_2/\xi_6,\,\,\,\,\,\, \xi_8=\sqrt{1-\xi_7^2},
\end{equation}
respectively, in order to obtain the multiple-domain SSC($2$) model after performing the Matryoshka sequence. 

\begin{figure}[h!]
    \includegraphics[width=\linewidth]{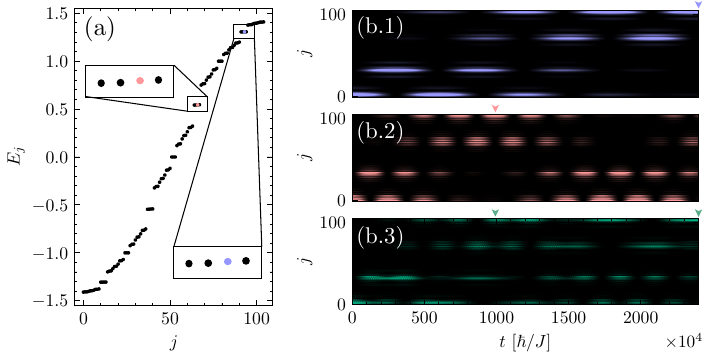}    
    \caption{(a) Energy spectrum of a three-domain SSC($3$) model with $\{\theta_1^{(3)},\theta_2^{(3)},\theta_3^{(3)},\theta_4^{(3)}\}=\{0.776012, 0.978122, 0.753036, 0.386992\}$, $t^{(2)}=0.704533$ and 3 unit cells per domain. Probability per site index $|\psi_j|^2$ over time of the three-domain SSC($3$) chain with initial state given by (b.1) the purple in-gap state, (b.2) the pink in-gap state and (b.3) the symmetric superposition of both (b.3), all truncated to the first domain. The colored arrows signal the probability maxima of the transfer processes.}
    \label{fig:ssc3transferdomains}
\end{figure}

Fig.\,\ref{fig:ssc3transferdomains}(a) presents the energy spectrum of a three-domain SSC($3$) model, where two domain wall states appear in each of its energy gaps, and Figs.\,\ref{fig:ssc3transferdomains}(b.1-b.3) display the transfer processes with initial states given by the purple in-gap state, pink in-gap state and  the symmetric superposition of both, respectively, truncated to the first domain. In contrast to what happens in the SSC($2$) case, part of the state that is being transferred is reflected by the domain walls and oscillates back and forth within each domain, forming the oscillating patterns seen in the plots. Nevertheless, domain wall mediated state transfer can still be achieved for both in-gap states. Consequently, for an initial state that combines in-gap states of distinct gaps, more than one transfer process occurs simultaneously. This is evidenced by the two transfer maxima signalled by the green arrows in Fig.\,\ref{fig:ssc3transferdomains}(b.3).

\section{Robustness against disorder}\label{sec:section5}
\begin{figure}[h!]
    \centering
    \includegraphics[width=\linewidth]{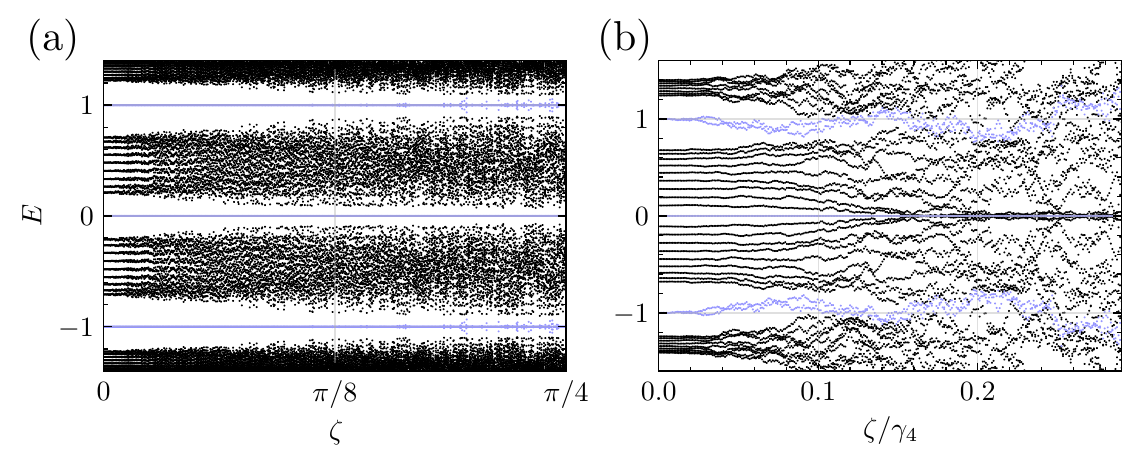}    
    \caption{Energy spectrum of a one domain SSC($2$) chain with $\theta_1^{(2)}=0.286479\pi$, $\theta_2^{(2)}=0.127324\pi$ and 15 unit cells as a function of the disorder strength $\zeta$ for (a) $\theta$-disorder and (b) uncorrelated off-diagonal disorder. The black points represent the energy values of the bulk states and the purple ones the edge states.}
    \label{disorder}
\end{figure}

The topological protection of the edge states is inherited by the root models from the parent chain \cite{PhysRevB.103.235425}. As such, the SSC($2$) edge states (with energies $\pm 1$) are protected against any type of disorder that preserves the sublattice symmetry of the parent SSH
chain \cite{viedma2024topological}. 
Let us consider angluar disorder ($\theta$-disorder) in a single domain SSC($2$) chain with $15$ unit cells, defined as $\theta_j^{(2)}\rightarrow \theta_j^{(2)}+\phi_j$, where $\phi_j$ is taken from a uniform distribution from the interval [$-\zeta/2$; $\zeta/2$], with $\zeta$ being the disorder strength. After squaring the $\theta$-disordered SSC($2$) model, the resulting parent SSH chain will have unitary onsite potentials that can be removed with an energy shift and will possess off-diagonal disorder, preserving chiral symmetry. 
This analysis establishes that the SSC($2$) model is robust against $\theta$-disorder [see Fig.~\ref{disorder}(a)] and fragile against uncorrelated off-diagonal disorder [see Fig.~\ref{disorder}(b)].
One can see from Fig.~\ref{disorder}(a) that, after a certain threshold value of the disorder strength $\zeta$, the disordered angle values may lead to some gap closing points and the disappearance of the finite-energy SSC($2$) edge states as indicated by the degeneracy lifting seen in some purple points with energies $
\pm 1$. Note that the zero-energy state remains immune to uncorrelated off-diagonal disorder, since it only has weight on $A$ and $C$ sites and is thus protected by the chiral symmetry of the SSC($2$) model.

\begin{figure}[h!]
    \centering
    \includegraphics[width=0.97\linewidth]{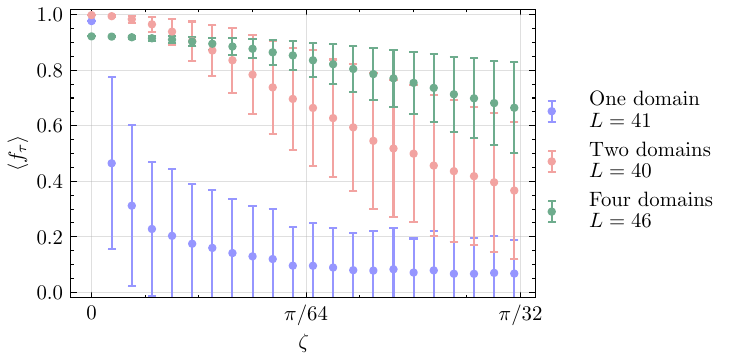}    
    \caption{Transfer fidelity of the general disordered SSC($2$) chain with $\theta_1^{(2)}=0.286479\pi$, $\theta_2^{(2)}=0.127324\pi$ and no state preparation as a function of the disorder strength $\zeta$ for the cases of one, two and four domains with length $L=41$, $L=40$ and $L=46$, respectively.  The dots (error bars) correspond to the average (standard deviation) over 1000 realizations.}
    \label{fidelity}
\end{figure}

Fig.~\ref{fidelity} shows the average fidelity of the positive energy left-edge state transfer as a function of disorder strength in a general disordered (diagonal disorder that follows the same uniform distribution in the interval [$-\zeta/2$; $\zeta/2$] and $\theta$-disorder) SSC($2$) chain with one, two and four domains without state preparation. In each of these cases, the fidelity values were obtained at the instant $\tau^{(n)}_d$, which is the transfer time of the disorder-free ($\zeta=0$) model. Since the diagonal disorder breaks sublattice symmetry and the transfer time depends directly on the energetic difference between the bonding and anti-bonding combinations of the states involved in the transfer, its fidelity decays for all cases.
However, as the number of domains increases, the effective hopping amplitudes between adjacent localized states increases, while the disorder strength gets comparatively smaller. This speeding up of the transfer process attenuates the effect of disorder, thereby enhancing its robustness (as evidenced by the diminishing error bars with increasing number of domains in Fig.~\ref{fidelity}). 
In the clean case, the one-domain SSC($2$) chain holds a lower fidelity than the two-domain SSC($2$) chain, which comes from the fact that the SSC($2$) transfer fidelity averages over the parent SSH and residual chain's fidelity values and, for an odd number of domains, the residual chain does not possess inversion symmetry (see Fig.~\ref{fig:squaringsscDdomains}), causing a slight asymmetry between the left and right edge states that negatively impacts the transfer fidelity.
After squaring the Hamiltonian of the SSC($2$) model with diagonal disorder expressed in the chiral basis,
\begin{equation}
\left(\mathcal{H}+\mathcal{V}\right)^2=\begin{pmatrix}
V^2+hh^\dagger&Vh+hU\\
h^\dagger V+Uh^\dagger& U^2+h^\dagger h
\end{pmatrix},
\label{matr}
\end{equation}
where $\mathcal{H}$ is given in Eq.\,(\ref{Hchiral}) and 
\begin{equation}
\mathcal{V}=\begin{pmatrix}
V&0\\
0&U
\end{pmatrix},
\end{equation}
with $V$ ($U$) containing the odd (even) disorder elements, one can see that it is no longer block diagonal due to these components coupling the parent and residual chains. Additionally, the $U^2$ term appearing in the right-hand side of Eq.\,(\ref{matr}) breaks the topologically protecting chiral symmetry of the parent SSH model. Both these effects explain the lack of protection of the boundary states of the SSC($2$) model against diagonal disorder.

\section{Conclusions} \label{sec:section6}

In this work, we studied the quantum state transfer process in high-root topological insulators. We have extended the approach of implementing multiple
domain walls along the Su-Schrieffer-Heeger chain \cite{platero} (to speed-up the transferring process
of the edge states exponentially) to the sine-cosine model \cite{Matryoshka}. Our analysis revealed that these models may find application in quantum computing and quantum telecommunications, since they offer multiple channels for state transfer that can be simultaneously activated. We have shown how the in-gap states involved in the state transfer are robust against some
types of angular disorder, and that robustness against general disorder can be enhanced by increasing the number of domains for a fixed chain size. This strategy can be used to mitigate the effects of information loss associated with quantum state transfer.
Sine-cosine models can be realized experimentally with photonic lattices \cite{wei2024realization,boross2019poor,perez2024transport}, where light injected in a waveguide has a probability of transitioning to their neighbouring guides depending on the distance between them and on the refraction index contrast with the embedding crystal.

\section*{Acknowledgments} 

This work was developed within the scope of Portuguese Institute for Nanostructures, Nanomodelling and Nanofabrication (i3N) Projects No.~UIDB/50025/2020, No.~UIDP/50025/2020, and No.~LA/P/0037/2020, financed by national funds through the Funda\c{c}\~{a}o para a Ci\^{e}ncia e Tecnologia (FCT) and the Minist\'{e}rio da Educa\c{c}\~{a}o e Ci\^{e}ncia (MEC) of Portugal. 
G.F.M. acknowledges financial support from FCT through the grant BI/UI96/12758/2025.
A.M.M. acknowledges financial support from i3N through the work Contract No.~CDL-CTTRI-91-SGRH/2024.

\section*{References}
\bibliography{references}


\end{document}